\documentstyle[12pt,epsfig]{article}
\def\lapproxeq{\lower .7ex\hbox{$\;\stackrel{\textstyle <}{\sim}\;$}}
\def\gapproxeq{\lower .7ex\hbox{$\;\stackrel{\textstyle >}{\sim}\;$}}

\newcommand{\be}{\begin{equation}}
\newcommand{\ee}{\end{equation}}

\begin{document}

\def\btop{\mathchar"1339}
\def\bmid{\mathchar"133D}
\def\bbot{\mathchar"133B}
\def\bs{\mathchar"1343}

\newfont{\sevenrm}{cmr7}
\newfont{\teni}{cmmi10}
\newfont{\seveni}{cmmi7}
\newfont{\sevensy}{cmsy7}
\newfont{\fiverm}{cmr5}
\newfont{\fivei}{cmmi5}
\newfont{\fivesy}{cmsy5}

\def\tenpoint{\normalbaselineskip=12pt
\abovedisplayskip 12pt plus 3pt minus 9pt
\belowdisplayskip 12pt plus 3pt minus 9pt
\abovedisplayshortskip 0pt plus 3pt
\belowdisplayshortskip 7pt plus 3pt minus 4pt
\smallskipamount=3pt plus1pt minus1pt
\medskipamount=6pt plus2pt minus2pt
\bigskipamount=12pt plus4pt minus 4pt
\def\rm{\fam0\tenrm}          \def\it{\fam\itfam\tenit}%
\def\sl{\fam\slfam\tensl}      \def\bf{\fam\bffam\tenbf}%
\def\smc{\tensmc}               \def\mit{\fam 1}%
\def\cal{\fam 2}%
\textfont0=\tenrm      \scriptfont0=\sevenrm    \scriptscriptfont0=\fiverm
\textfont1=\teni       \scriptfont1=\seveni     \scriptscriptfont1=\fivei
\textfont2=\tensy      \scriptfont2=\sevensy    \scriptscriptfont2=\fivesy
\textfont3=\tenex      \scriptfont3=\tenex      \scriptscriptfont3=\tenex
\normalbaselines\rm}

\hfill DTP/94/32
\begin{center}
{\Large{\bf Structure functions and small $x$ physics}}$^\dagger$
\end{center}

\vspace*{.5cm}
\begin{center}
A.D.\ Martin
\end{center}

\begin{center}
Department of Physics, University of Durham, DH1 3LE, England.
\end{center}

\vspace*{.5cm}

\begin{abstract}
We review recent developments in the determination of parton densities from
deep
inelastic and related data.  We show how the asymmetries observed in the
$W^{\pm}$ rapidity distributions and in $pp/pn$ Drell-Yan production further
constrain the partons at moderate $x$.  We compare the GLAP and BFKL
descriptions of the recent measurements of $F_2(x,Q^2)$ at HERA.  We survey
processes which may be used to identify BFKL dynamics.
\end{abstract}

\vspace*{1.2cm}

\noindent {\large {\bf 1.  Introduction}}
\vspace*{0.5cm}

An idea of the rapid improvement in our knowledge of the structure function,
$F_2(x,Q^2)$, of deep inelastic electron (or muon)-proton scattering can be
glimpsed from the data shown in Fig.\ 1.  In fact it should not be too long
before we have precision measurements (perhaps as good as $\pm 5$\%) of $F_2$
at
HERA almost down to $x \sim 10^{-4}$.  The data show a dramatic rise of $F_2$
with decreasing $x$, which some optimists had anticipated might occur from the
precocious onset of the leading $\alpha_s$log$(1/x)$ resummation of soft gluon
emissions.

The curves marked D$^{\prime}_-$ and D$^{\prime}_0$ are simply two
extrapolations obtained from parton fits to the fixed target data, which at the
time it was thought might span the subsequent measurements of $F_2$ at HERA --
one represents the upper limit expected for the precocious resummation
behaviour, the other the lower limit of conventional (Regge-motivated)
expectations.  Such extrapolations are notoriously unreliable, and have failed
in the past, much to the glee of NMC.

Section 2 is mainly devoted to recent developments in the determination of
partons in the $x \gapproxeq 10^{-2}$ region.  For the purposes of illustration
we mainly concentrate on the MRS analyses.  In Section 3 we discuss the
description of $F^{ep}_2$ at small $x$ and find that both GLAP-based
extrapolations and BFKL-based predictions have sufficient freedom to describe
the HERA data.  Section 4 surveys some theoretical studies of processes which
might be able to distinguish between GLAP and BFKL dynamics at small $x$.

\vfil
\noindent ---------------\hfil

\noindent $^\dagger$ Invited talk at the International Workshop on Deep
Inelastic Scattering and related subjects, Eilat, Israel, February 6-11th,
1994.

\newpage

{\protect{\tenpoint
\begin{figure}[h]
\begin{center}
{}~ \epsfig{file=/pdg/rachel/adm/fig1.ps,width=200pt}
\end{center}
\caption{{\protect{\tenpoint The $x$ dependence of the structure function
$F_2(x,Q^2)$ at $Q^2
= 15$ GeV$^2$.  The data are from the BCDMS [1], NMC [2],
H1 [3] and ZEUS [4] collaborations.  The curves are calculated
from the MRS(D$^{\prime}_0$,D$^{\prime}_-$) sets of partons [5].}}
\label{1}
}
\end{figure}
}}

\vspace*{.5cm}

\noindent {\large {\bf 2.  Recent developments in the determination of parton
densities}}
\vspace*{0.5cm}

There is a long history of determining parton distributions from deep inelastic
scattering and related data.  The increased precision in the experimental
measurements over the last couple of years has led to a considerable
improvement
in our knowledge of the partons, at least for $x \gapproxeq 0.01$.  The recent
experiments are listed in Table 1, together with the leading order partonic
subprocess, which help us to see which partons are constrained by the various
processes.  In particular it is useful to note that the four deep-inelastic
structure functions can, to leading order, be expressed in terms of parton
densities in the form
\begin{eqnarray}
F^{\mu p}_2 - F^{\mu n}_2 & = & {\textstyle \frac{1}{3}}x(u + \bar{u} - d -
\bar{d}) \\
{\textstyle \frac{1}{2}}(F^{\mu p}_2 + F^{\mu n}_2) & = & {\textstyle
\frac{5}{18}}x(u + \bar{u} + d + \bar{d} + {\textstyle \frac{4}{5}}s) \\
F^{\nu N}_2 = F^{\bar{\nu}N}_2 & = & x(u + \bar{u} + d + \bar{d} + 2s) \\
{\textstyle \frac{1}{2}}x(F^{\nu N}_3 + F^{\bar{\nu}N}_3) & = & x(u - \bar{u} +
d - \bar{d})
\end{eqnarray}
where $N$ is an isoscalar nuclear target.  We have included the $s$ quark (and
\linebreak
\newpage

\noindent assumed\footnote{In principle $s$ and $\bar{s}$ could be
distinguished
by accurate $\nu N$ and $\bar{\nu}N$ data.} that $s = \bar{s}$), but neglected
the small $c$ quark contribution.  These four observables, $F_i(x,Q^2)$,
measured in the fixed-target $(\mu N \rightarrow \mu X$ and $\nu N \rightarrow
\mu X)$ deep-inelastic experiments, do not on their own determine the five
distributions  $u, d, \bar{u}, \bar{d}$ and $s$, but only four combinations,
which we may take to be $u + \bar{u}, \, d + \bar{d}, \, \bar{u} + \bar{d}$ and
$s$.  The difference $\bar{d}-\bar{u}$ is not determined, and, moreover, at
leading order it appears that the only constraint on the gluon is through the
momentum sum rule.  In practice the situation is worse.  From (2) and (3) we
see
that the strange quark distribution is essentially determined by the structure
function difference
\begin{equation}
xs(x) \; \simeq \; {\textstyle \frac{5}{6}} F^{\nu N}_2(x) - 3F_2^{\mu D}(x)
\end{equation}
which is sensitive to the relative normalization of the two data sets and to
heavy target corrections to the neutrino measurements.  However, we see from
Table 1 that valuable independent information comes from other processes: (i)
the gluon is constrained by prompt photon production, where $g(x)$ enters to
leading order, (ii) the strange sea is measured by neutrino-induced dimuon
production (see Fig.\ 2), and (iii) there is a measurement of the difference
$\bar{d}-\bar{u}$, which was released for the first time at this conference,
from the observed asymmetry of Drell-Yan production in $pp$ and $pn$ collisions
(see subsection (c)).

\begin{table}

\begin{itemize}
{\protect {\tenpoint
\item[Table 1] The experimental data used to determine the MRS parton
distributions.  The last column gives an indication of the main type of
constraint imposed by a particular set of data.}
}
\end{itemize}

\vspace*{1cm}

\begin{tabular}{|l|l|l|}    \hline
                   &                           &                         \\
{\bf Process/}      &     {\bf Leading order}    & {\bf Parton determination}
\\
{\bf Experiment}   &  {\bf subprocess}          &                    \\
&                            &                  \\  \hline
&\hfill \raisebox{-0.5ex}[0.5ex][0.5ex]{$\btop$}&                      \\
{\bf DIS} $\mbox{\boldmath $(\mu N \rightarrow \mu X)$}$ &  $\gamma^*q
\rightarrow q$ \hfill {\arrayrulewidth=1pt\vline}\hspace*{4pt}& Four structure
functions $\rightarrow$ \\
BCDMS, NMC& \hfill {\arrayrulewidth=1pt\vline}\hspace*{4pt}& \hspace*{1cm}  $u
+ \bar{u}$  \\
$F^{\mu p}_2,F^{\mu n}_2$& \hfill {\arrayrulewidth=1pt\vline}\hspace*{4pt}&
\hspace*{1cm} $d + \bar{d}$   \\
      &\hfill $\bmid$ & \hspace*{1cm}  $\bar{u} + \bar{d}$  \\
{\bf DIS} $\mbox{\boldmath $(\nu N \rightarrow \mu X)$}$ & $W^*q \rightarrow
q^{\prime}$  \hfill {\arrayrulewidth=1pt\vline}\hspace*{4pt}& \hspace*{1cm}
$s$ (assumed = $\bar{s}$),\\
CCFR (CDHSW)    &\hfill {\arrayrulewidth=1pt\vline}\hspace*{4pt}& but only
$\int xg(x)dx \simeq 0.5$ \\
$F^{\nu N}_2,xF^{\nu N}_3$    &\hfill {\arrayrulewidth=1pt\vline}\hspace*{4pt}&
[$\bar{u}-\bar{d}$ is not determined] \\
                &\hfill \raisebox{0.5ex}[1.5ex][1.5ex]{$\bbot$} &
                     \\
$\mbox{\boldmath $\nu N \rightarrow \mu^+\mu^-X$}$ &  $\nu s \rightarrow \mu^-
c$    & $s \approx \frac{1}{2}\bar{u}$ (or $\frac{1}{2}\bar{d}$) \\
CCFR            &   $\;\;\;\;\;\;\;\;\;\;\;\;\;\;\hookrightarrow \mu^+$     &
\\
               &                            &                           \\
{\bf DIS (HERA})    &   $\gamma^*q \rightarrow q$   &   $\lambda$     \\
$F^{ep}_2$ (H1,ZEUS)   &                  & $(x\bar{q} \sim xg \sim
x^{-\lambda}$, via $g \rightarrow q\bar{q})$    \\
                      &                         &                     \\
$\mbox{\boldmath $pp \rightarrow \gamma X$}$  &  $qg \rightarrow \gamma q$  &
$g(x \approx 0.4)$ \\
WA70 (UA6)    &                  &                        \\
              &                   &                          \\
$\mbox{\boldmath $pN \rightarrow \mu^+\mu^- X$}$   &  $q\bar{q} \rightarrow
\gamma^*$  &  $\bar{q} = ...(1-x)^{\eta_S}$ \\
E605           &                     &                             \\
               &                     &                                 \\
$\mbox{\boldmath $pp, pn \rightarrow \mu^+\mu^- X$}$ & $u\bar{u},d\bar{d}
\rightarrow \gamma^*$   & $(\bar{u}-\bar{d})$ at $x = 0.18$  \\
NA51             &  $u\bar{d},d\bar{u} \rightarrow \gamma^*$   &     \\
                 &                   &                               \\
$\mbox{\boldmath $p\overline{p} \rightarrow WX(ZX)$}$    &  $ud \rightarrow W$
&  $u,d$ at
$x_1x_2s \simeq M^2_W \rightarrow$  \\
UA2, CDF, D0          &             & \hspace*{1cm}  $x \approx 0.13$  CERN  \\
                       &             & \hspace*{1cm} $x \approx 0.05$ FNAL  \\
$\;\;\;\;\;\mbox{\boldmath $\rightarrow W^{\pm}$}$ {\bf asym}   &         &
slope of $u/d$ at $x \approx 0.05$  \\
$\;\;\;\;\;\;\;\;\;\;$ CDF            &              &              \\
                         &                         &          \\  \hline
\end{tabular}

\end{table}

\vspace*{.5cm}

\noindent {\bf (a) MRS global analysis}

Here we describe the MRS determination \cite{MRSD} of the parton densities,
$f_i(x,Q^2)$, from a global analysis of the data, which was performed before
the
HERA measurements of $F^{ep}_2$ became available.  Only deep inelastic data
with
$Q^2 > 5$ GeV$^2$ (and $W^2 > 10$ GeV$^2$) were used.  The input distributions
at $Q^2_0 = 4$ GeV$^2$ were parametrized in the form
\begin{equation}
xf_i(x,Q^2_0) \; = \; A_i x^{-\lambda_i}(1-x)^{\beta_i} (1 + \gamma_i
x^{\frac{1}{2}} + \delta_i x)
\end{equation}
for $i =$ the valence quarks $u_{{\rm v}}, d_{{\rm v}}$, the sea $S$, and the
gluon $g$.   Three of the four $A_i$'s are determined by the momentum and
flavour sum rules.  Since, pre-HERA, there were no data for $x < 10^{-2}$, we
imposed two different extrapolations of the gluon and sea quark distributions
to
small $x$ (with the expectation that they should span the forthcoming HERA
measurements).  To be precise for the gluon and sea quarks we choose
\begin{displaymath}
\lambda_i \; \equiv \; 0 \;\;\; {\rm for \; the \; D}^{\prime}_0 \; {\rm set}
\end{displaymath}
\begin{equation}
\lambda_i \; \equiv \; 0.5 \;\;\; {\rm for \; the \; D}^{\prime}_- \; {\rm set}
\end{equation}
where, since $\sigma(\gamma^*p) \sim s^{\lambda_i}$, the first choice
corresponds to conventional Pomeron exchange leading to constant total cross
sections, and the second is motivated by the small $x$ behaviour expected from
the BFKL (Balitsky, Fadin, Kuraev, Lipatov) equation \cite{BFKL}.  The
$\beta_i$
and the valence $\lambda_i$ are left as free parameters although we have some
idea what values will be physically reasonable.  First, the small $x$ behaviour
\linebreak

\newpage

\noindent  anticipated from meson Regge exchanges for the valence quark
distributions would correspond to $\lambda_{{\rm v}} \sim -0.5$.  Secondly,
naive counting rule arguments imply that $xf_i \sim (1-x)^{2n-1}$ as $x
\rightarrow 1$, where $n$ is the minimum number of spectator quarks which
accompany the struck parton.  This rule would suggest $\beta_i \sim$ 5, 3 and 7
for, respectively, the gluon, valence and sea quark distributions.

In the MRS analysis the flavour structure of the quark sea $S$ at $Q^2_0 = 4$
GeV$^2$ is taken to be
\begin{eqnarray}
2 \bar{u} & = & 0.4S - \Delta   \nonumber  \\
2 \bar{d} & = & 0.4S + \Delta    \nonumber  \\
2\bar{s}  & = & 0.2S  \nonumber  \\
x\Delta & \equiv & x(\bar{d}-\bar{u}) = A_{\Delta} x^{-\lambda_{\rm
v}}(1-x)^{\eta_S} .
\end{eqnarray}
As mentioned before, the most reliable method to estimate the strength of the
strange sea distribution $\bar{s}(x,Q^2)$ is to observe deep-inelastic dimuon
production, $\nu N \rightarrow \mu^-\mu^+ X$, for which the dominant subprocess
is $\nu s \rightarrow \mu^- + c(\rightarrow \mu^+)$.  Early dimuon data were
the
motivation of the 50\% suppression of the input strange sea in the MRS
analyses,
which was assumed to be independent of $x$.  Recently this input assumption has
been checked by a next-to-leading order analysis performed by the CCFR
collaboration \cite{CCFR2} on their dimuon data.  The MRS assumption is in
excellent agreement with their results (see Fig.\ 2) and moreover CCFR find
that
the ratio $\bar{s}/(\bar{u}+\bar{d})$ is essentially independent of $x$.  It
has
been emphasized \cite{GEN} that the strange sea as measured in neutrino
scattering  should be different from that in muon scattering on account of the
different mass thresholds in $W^*g \rightarrow s\bar{c}$ versus $\gamma^* g
\rightarrow s\bar{s}$.  In practice the neutrino data have been corrected by
CCFR to approximately take into account the $m_c \neq 0$ effects and allow for
this difference.

The first evidence that $\bar{d} \neq \bar{u}$, and hence the need to introduce
$\Delta$ in (8), came from the NMC study of the Gottfried sum rule \cite{NMC3}.
We return to discuss this later.  Finally, the heavy quark contributions are
determined by assuming that
\begin{equation}
\bar{Q} = 0 \hspace*{.5cm} {\rm for} \hspace*{.5cm} Q^2 \leq 4 m^2_Q
\hspace*{1cm} {\rm with} \hspace*{.5cm} Q = c,b,...,
\end{equation}
and by generating non-zero distributions at higher $Q^2$, via $g \rightarrow
Q\bar{Q}$, using QCD evolution with $m_Q = 0$.  Near threshold clearly this
prescription is unreliable, but it has been shown to give reasonable results at
higher $Q^2$ \cite{MRSC}.  The charm quark distribution in the proton has
attracted much attention \cite{CHARM} but the phenomenological situation is far
from clear.

The input distributions (6) and (8) are evolved up in $Q^2$ using
next-to-leading order GLAP or Altarelli-Parisi equations \cite{GLAP} and fitted
to deep-inelastic structure functions (with $Q^2 > 5$ GeV$^2$ and $W^2 > 10$
GeV$^2$) and related data.  The distributions are defined in the
$\overline{{\rm
MS}}$ renormalization and mass factorization scheme.  The fitted value of the
QCD scale parameter is $\Lambda_{\overline{{\rm MS}}}(n_f = 4) = 230$ MeV (that
is $\alpha_s(M^2_Z) = 0.112_5)$, which is consistent with the values $\alpha_s
=$ 0.113 and 0.111 found by analyses of BCDMS \cite{MV} and CCFR \cite{CCFRA}
subsets of the overall data.

\newpage

{\protect{\tenpoint
\begin{figure}[h]
\begin{center}
{}~ \epsfig{file=/pdg/rachel/adm/fig2.ps,width=200pt}
\end{center}
\caption{{\protect{\tenpoint The shaded band shows the strange quark
distribution extracted by CCFR from their dimuon data [7], together
with the MRS distributions of ref.\ [5].}}
\label{2}
}
\end{figure}
}}

{\protect{\tenpoint
\begin{figure}[h]
\begin{center}
{}~ \epsfig{file=/pdg/rachel/adm/fig3.ps,width=200pt}
\end{center}
\caption{{\protect{\tenpoint The description of the NMC data [2]
for the structure function ratio $F^{\mu n}_2/F^{\mu p}_2$ given by the
MRS partons of ref.\ [5].}}
\label{3}
}
\end{figure}
}}

{\protect{\tenpoint
\begin{figure}[p]
\begin{center}
{}~ \epsfig{file=/pdg/rachel/adm/fig4.ps,width=150pt}
\end{center}
\caption{{\protect{\tenpoint The BCDMS [1] and NMC [2] measurements of
$F^{\mu p}_2$, together with the MRS fit [5].}}
\label{4}
}
\end{figure}
}}

{\protect{\tenpoint
\begin{figure}[t]
\begin{center}
{}~ \epsfig{file=/pdg/rachel/adm/fig5.ps,width=200pt}
\end{center}
\caption{{\protect{\tenpoint The continuous curves show the description of
the CCFR [14] measurements of $F^{\nu N}_2 (x,Q^2)$ and $xF^{\nu N}_3 (x,Q^2)$
by the MRS(D$^{\prime}_0)$ set of partons [5].  The data are shown after
correction for the heavy target effects and after the overall renormalization
of 0.95 required by the global fit.}}
\label{5}
}
\end{figure}
}}

Figs.\ 3-5 show the high precision of recent deep-inelastic structure function
measurements, as well as the quality of the fit.  Accurate data exist also for
related processes and are equally well described.  This should mean that the
parton distributions are well determined, at least in the $x$ range $0.02
\lapproxeq x \lapproxeq 0.7$ where data are available for a full range of
different processes.   At smaller $x$, $x \lapproxeq 10^{-3}$, the measurements
of $F^{ep}_2$ at HERA give information on the $x$ behaviour of the quark sea.
We will discuss the expectations of the small $x$ behaviour in some detail
later.  For the moment we simply note that it is possible to \lq tune' the
exponent $\lambda_i$ of the input sea and gluon distributions of (6) to obtain
the best global fit to a data set which now includes the new HERA measurements
of $F_2$ \cite{H1,ZEUS}.  The result, set MRS(H) of partons \cite{MRSH}, has
$\lambda_i = 0.3$.

Before turning to a theoretical discussion of the small $x$ region, we must
mention two recent measurements which impose valuable new constraints on the
partons in the $x \gapproxeq 0.02$ region.  The first is the measurement of the
rapidity asymmetry of $W^{\pm}$ bosons produced in $p\bar{p}$ collisions and
the
second is the Drell-Yan asymmetry observed in $pp$ and $pn$ collisions.

\vspace*{.5cm}

\noindent {\bf (b) $\mbox{\boldmath $W^{\pm}$}$ rapidity asymmetry}

$W^{\pm}$ bosons produced in $\bar{p}p$ collisions have different rapidity
$(y_W)$ distributions.  A comparison of parton distributions shows that the $u$
quark in the proton tends to go more forward than the $d$ quark, so that the
$W^+$ is produced preferentially in the direction of the incident proton, while
the $W^-$ tends to go in the direction of the antiproton.  Thus we anticipate
that the asymmetry
\begin{equation}
A_W(y_W) \; = \; \frac{\sigma (W^+)-\sigma (W^-)}{\sigma (W^+) + \sigma (W^-)}
\end{equation}
will be positive and increase with the rapidity $y_W$ of the produced $W^{\pm}$
bosons.  It turns out that the asymmetry $A_W$ is correlated with the slope of
$F^n_2/F^p_2$ in the  $x \simeq M^2_W/\sqrt{s}$ region \cite{MRSW,BODEK} --
larger slopes tend to imply greater asymmetry.  In practice it is the rapidity
asymmetry of the $\ell^{\pm}$ decay leptons which is observed
\begin{equation}
A(y_{\ell}) \; = \; \frac{\sigma(\ell^+) - \sigma(\ell^-)}{\sigma(\ell^+) +
\sigma(\ell^-)} .
\end{equation}
Folding in the $W \rightarrow \ell \nu$ decay dilutes the asymmetry, but the
measurements \cite{BODEK} shown in Fig.\ 6 indicate that it is still sizeable.
The present data favour the recent MRS rather than the CTEQ parton sets.   As
the experimental precision increases we see that there is a great potential for
the asymmetry to further discriminate between parton sets -- indeed these data
should be included in future global analyses from the beginning.

\vspace*{.5cm}
\noindent {\bf (c) $\bar{d}-\bar{u}$ and the Drell-Yan asymmetry}

Inspection of the leading order expressions (1)-(4) shows that measurement of
the four independent deep-inelastic structure functions do not determine
$\bar{d}-\bar{u}$ on a $(x,Q^2)$ point-by-point basis.  The difference
$\bar{d}-\bar{u}$ is only constrained if the form of the $x$ dependence of
partons is assumed.  Indeed prior to 1992 all global analyses set $\bar{u} =
\bar{d}$.  The first hint that $\bar{u} \neq \bar{d}$ came from an NMC study
\cite{NMC1} of the Gottfried sum rule.  NMC found that their data gave
\begin{displaymath}
\int^{0.8}_{0.004} \frac{dx}{x} (F^{\mu p}_2 - F^{\mu n}_2) \; = \; 0.227 \pm
0.007 ({\rm stat.}) \pm 0.014 ({\rm sys.})
\end{displaymath}
at $Q^2 = 4$ GeV$^2$, as compared to the Gottfried sum
\begin{displaymath}
I_{{\rm GSR}} \; = \; \int^1_0 \frac{dx}{x}(F^{\mu p}_2 - F^{\mu n}_2) \; = \;
{\textstyle \frac{1}{3}} \int^1_0 dx(u_{\rm v}-d_{\rm v}) + {\textstyle
\frac{2}{3}} \int^1_0 dx(\bar{u}-\bar{d})
\end{displaymath}
\begin{equation}
= \; {\textstyle \frac{1}{3}} \hspace*{1cm} {\rm if} \hspace*{.5cm} \bar{u} =
\bar{d} .
\end{equation}

\newpage

{\protect{\tenpoint
\begin{figure}[h]
\begin{center}
{}~ \epsfig{file=/pdg/rachel/adm/fig6.ps,width=200pt}
\end{center}
\caption{{\protect{\tenpoint A plot of the CDF data for the lepton charge
asymmetry, (11), compared to the predictions of various sets of partons
[5, 18, 19]. The plot is taken from ref.\ [16]. These data were fitted in
the MRS(A) analysis [20] which gives a description essentially equal to that
of MRS(D$^{\prime}_-$).}}
\label{6}
}
\end{figure}
}}

{\protect{\tenpoint
\begin{figure}[h]
\begin{center}
{}~ \epsfig{file=/pdg/rachel/adm/fig7.ps,width=200pt}
\end{center}
\caption{{\protect{\tenpoint The structure function difference
$F^{\mu p}_2-F^{\mu n}_2$ measured by NMC [22] compared with the description
by partons from refs.\ [5, 18].}}
\label{7}
}
\end{figure}
}}

\newpage

\noindent Using their data to extrapolate over the unmeasured regions of $x$,
NMC obtained
$I_{{\rm GSR}} = 0.240 \pm 0.016$ \cite{NMC1}.\footnote{An updated analysis of
the Gottfried sum by NMC \cite{NMC1} gives $I_{{\rm GSR}} = 0.258 \pm 0.010$
(stat.) $\pm 0.015$ (sys.) at $Q^2 = 4$ GeV$^2$.  If $F^{\mu n}_2$ is corrected
for deuteron screening effects then $I_{{\rm GSR}}$ is reduced to $\simeq
0.23$.}  The most plausible explanation of the difference between the NMC value
and the Gottfried sum of $\frac{1}{3}$  is that $\bar{d} > \bar{u}$.  In the
MRS
\lq D' fits $\bar{d}-\bar{u}$ is parametrized according to (8) with the
parameter $A_{\Delta}$ chosen to reproduce the measured value of $I_{{\rm
GSR}}$.  It is interesting to note, however, that even including the NMC data,
it is still possible to maintain $\bar{u} = \bar{d}$ and obtain an equally good
global description of the data, but at the expense of a contrived small $x$
behaviour of the valence distributions, as in set S$^{\prime}_0$ of ref.\
\cite{MRSD}.  The comparison of the $(F^p_2-F^n_2)$ NMC data with
S$^{\prime}_0$
and D$^{\prime}_0$ is shown in Fig.\ 7 -- the area under the curves gives
$I_{{\rm GSR}} = 0.333$ and $0.256$ respectively, the difference coming
primarily from the unmeasured  $x \lapproxeq 10^{-2}$ region.  The most
physical
assumption is to allow $\bar{u} \neq \bar{d}$, as in the \lq D'-type fits.
Nevertheless the conclusion is that the detailed structure of $\bar{d}-\bar{u}$
is not determined by the available deep-inelastic data.

As the $W^{\pm}$ rapidity asymmetry measurements improve they become
increasingly sensitive to the behaviour of $\bar{u}$ and $\bar{d}$ in the
region
$x \simeq M^2_W/\sqrt{s}$.  However a more direct method \cite{ES} to obtain
information on $\bar{d}-\bar{u}$ is to compare (Drell-Yan) lepton-pair
production in $pp$ and $pn$ collisions, via the asymmetry
\begin{equation}
A_{{\rm DY}} \; = \; \frac{\sigma_{pp}-\sigma_{pn}}{\sigma_{pp}+\sigma_{pn}} .
\end{equation}
Because $u > d$ in the proton, the asymmetry $A_{{\rm DY}}$ is positive for
parton sets with $\bar{d}-\bar{u}$ zero or small, but $A_{{\rm DY}}$ is reduced
and can even be negative for parton sets with larger $\bar{d}-\bar{u}$
\cite{MRSDY}.  An asymmetry measurement by the NA51 collaboration \cite{NA51},
\begin{equation}
A_{{\rm DY}} \; = \; -0.09 \pm 0.02 \pm 0.02
\end{equation}
at $x = 0.18$, was announced at this conference.  This result, which implies
$\bar{u}/\bar{d} \simeq 0.5$ at $x = 0.18$, corresponds to a breaking of
flavour
symmetry considerably beyond that associated with the
MRS(D$^{\prime}_0$,D$^{\prime}_-$) partons and almost as large as that
calculated using CTEQ2M partons, see Fig.\ 8.

We may conclude that, as far as the asymmetries are concerned, the MRS partons
accurately reproduce the $W^{\pm}$ rapidity asymmetry (Fig.\ 6) but do not
describe the new Drell-Yan asymmetry measurement.  On the other hand the CTEQ
partons describe the $W^{\pm}$ asymmetry rather poorly, but offer a much better
description of the Drell-Yan asymmetry.

\vspace*{.5cm}

\noindent {\bf (d) Present status of parton distributions}

We noticed that the high precision deep-inelastic data are well described by
the
parton distributions but that they do not pin down the $\bar{d}-\bar{u}$
combination.  However\linebreak

\newpage

{\protect{\tenpoint
\begin{figure}[h]
\begin{center}
{}~ \epsfig{file=/pdg/rachel/adm/fig8.ps,width=200pt}
\end{center}
\caption{{\protect{\tenpoint The $pp/pn$ Drell-Yan asymmetry of (13) measured
by NA51 [25] compared to values obtained by various sets of partons
[5, 15, 20, 18].  The data point was included in the MRS(A) analysis [20].}}
\label{8}
}
\end{figure}
}}

{\protect{\tenpoint
\begin{figure}[h]
\begin{center}
{}~ \epsfig{file=/pdg/rachel/adm/fig9.ps,width=200pt}
\end{center}
\caption{{\protect{\tenpoint The MRS(A) [20] and MRS(H) [15] parton
distributions at $Q^2 = 20$ GeV$^2$ shown by continuous and dotted curves
respectively.  Unlike MRS(H), MRS(A) incorporated constraints from the
asymmetry
data of Figs.\ 6 and 8 in the analysis.}}
\label{9}
}
\end{figure}
}}

\newpage
\noindent other measurements give independent information on the
partons.  In particular the new asymmetry measurements probe fine details of
the
quark distributions.

Can we use the residual freedom of $\bar{d}-\bar{u}$ to modify the MRS(H) set
of
partons so as to describe the new Drell-Yan asymmetry measurement \cite{NA51}
while maintaining the quality of the global fit to the data?  In particular can
we maintain the quality of the description of the $W^{\pm}$ asymmetry (Fig.\ 6)
and of the accurate NMC measurements of $F^n_2/F^p_2$?  It turns out that such
a
fit is possible -- we denote the new set of partons by MRS(A) \cite{MRSA}.  The
asymmetries resulting from this set of partons are shown in Figs.\ 6 and 8.  We
would expect the MRS(A) partons to be very similar to those of MRS(H), except
that $\bar{d}-\bar{u}$ would be much larger while at the same time
approximately
conserving $\bar{u}+\bar{d}$, $d+\bar{d}$ and $u+\bar{u}$ (as required by
(1)-(4)).  Thus we anticipate an increase of $\bar{d}$ which is compensated by
a
corresponding decrease in $\bar{u}$ and $d$, which in turn requires a similar
increase in $u$.  The comparison of MRS(H) and MRS(A) parton sets is shown in
Fig.\ 9, and displays these trends in the region $0.02 \lapproxeq x \lapproxeq
0.7$ where a full set of deep inelastic data exist.

Although the quark distributions are well-constrained by the data, the same is
not so true for the gluon.  Within the global analyses, we can identify three
types of constraint on the gluon: (i) the momentum sum rule, (ii) the prompt
photon production measurements and (iii) the observed overall pattern of scale
violations, $\partial F_i/\partial$log$Q^2$, for $x \lapproxeq 0.1$.  The
measurements of $F^{ep}_2$ at HERA probe primarily the behaviour of the sea
quark distributions at small $x$, although accurate measurements of $\partial
F^{ep}_2/\partial$log$Q^2$ will further constrain the gluon.  If the observed
steep rise of $F_2$ with decreasing $x$ is attributed to the Lipatov or BFKL
perturbative QCD mechanism via $g \rightarrow q\bar{q}$, then the same
behaviour
should be seen also in the gluon distribution.  A direct measurement of the
gluon in the $x \lapproxeq 10^{-2}$ region would be invaluable.

In addition to the potential constraints coming from accurate measurements of
$\partial F_2/\partial$log$Q^2$, there are several other possible ways of
measuring the gluon at HERA, including $F_L$, $J/\psi$, $Q\bar{Q}$...
production.  In principle the observation of the longitudinal structure
function
is a gold-plated measurement, but in practice it will be difficult.  To extract
$F_L$ as a function of $x,Q^2$, requires observation of the $y = Q^2/xs$
dependence of deep inelastic scattering, which means having electron-proton
collisions at different c.m.\ energies $\sqrt{s}$.

{\protect{\tenpoint
\begin{figure}[p]
\begin{center}
{}~ \epsfig{file=/pdg/rachel/adm/fig10.ps,width=300pt}
\end{center}
\caption{{\protect{\tenpoint The same-side/opposite-side dijet ratio predicted
by parton sets of refs.\ [5, 15] compared with preliminary,
uncorrected CDF data.}}
\label{10}
}
\end{figure}
}}

There are also possibilities to obtain information on the gluon from hadron
colliders, in particular the Fermilab $p\bar{p}$ collider.  At $\sqrt{s} = 1.8$
TeV, $x$ values comparable to those currently measured at HERA can be probed
either by observing small mass systems produced centrally (e.g. $b\bar{b}$ or
Drell-Yan production \cite{BODEK}) or by more massive final states at large
rapidity $y$ (e.g. same-side $W$+jet or dijet production).  As an illustration
Fig.\ 10 shows the ratio of same-side to opposite-side jet cross sections from
uncorrected CDF data, as a function of the equal and opposite rapidities $y$.
At large $y$ the same-side cross section originates dominantly from
subprocesses
of  $g$(small $x$)$q_{{\rm val}}$(large $x$) origin.  The curves are the
MRS(D$^{\prime}_-$,D$^{\prime}_0$,H) predictions evaluated at leading order
with
a renormalization/factorization scale chosen to mimic the
\linebreak

{\protect{\tenpoint
\begin{figure}[t]
\begin{center}
{}~ \epsfig{file=/pdg/rachel/adm/fig11.ps,width=200pt}
\end{center}
\caption{{\protect{\tenpoint (a) The ladder diagram arising in the GLAP
evolution of the (non-singlet) quark distribution.  The unintegrated gluon
distribution, $f(x,k^2_T)$, is obtained from the BFKL sum of leading
log$(1/x)$ \lq\lq ladder" diagrams of the type shown in (b).}}
\label{11}
}
\end{figure}
}}

\noindent next-to-leading order
corrections.  Further details of this analysis can be found in ref.\
\cite{MRSDJ}.  We see that at large rapidity the ratio is sensitive to the
small
$x$ behaviour of the gluon and that the preliminary CDF data favour the
singular
distributions.  Dijet production from \lq direct' photons at HERA offers
similar
possibilities of probing the gluon at small $x$ via $\gamma g \rightarrow
q\bar{q}$  \cite{FR}.

\vspace*{1cm}

\noindent {\large {\bf  3.  Small $x$ dynamics and $F_2(x,Q^2)$}}
\vspace*{.5cm}

The parton analyses use Altarelli-Parisi (or GLAP) equations to evolve up in
$Q^2$ from a set of starting distributions at $Q^2_0$.  At lowest order, these
evolution equations effectively resum the leading order $\alpha_s$log$Q^2$
contributions which, in an axial gauge, amounts to a summation of ladder
diagrams of the type shown in Fig.\ 11(a).  The $(\alpha_s$log$Q^2)^n$
contribution comes from the diagram with $n$ rungs with the transverse momenta
strongly ordered along the ladder, that is $Q^2 \gg k^2_{nT} \gg ... \gg
k^2_{1T}$. At small $x$ we encounter large log$(1/x)$ terms which have to be
resummed.  Indeed the dramatic rise observed in $F_2$ with decreasing $x$ may
be
associated with the growth of the gluon density which arises from the BFKL
resummation of these terms; a growth which, via $g \rightarrow q\bar{q}$, is
transmitted to the sea quarks probed by the photon.  But first we discuss
gluons
obtained using GLAP evolution.

\vspace*{.5cm}

\noindent {\bf (a) GLAP expectations}

What does Altarelli-Parisi evolution say about the small $x$ behaviour of the
gluon?  Suppose we evolve from the non-singular gluon form $xg(x,Q^2_0)
\rightarrow$ constant as $x \rightarrow 0$, then Altarelli-Parisi evolution in
$Q^2$ generates an increasingly steep behaviour in $x$.\linebreak

\newpage

\noindent Indeed at small $x$ and
large $Q^2$ it is straightforward to show that
\begin{equation}
xg(x,Q^2) \sim {\rm exp}\left( 2\left[\xi (Q^2_0,Q^2){\rm
log}(1/x)\right]^{\frac{1}{2}}\right)
\end{equation}
up to a slowly varying function of the argument of the exponential, where the
\lq\lq evolution length"
\begin{displaymath}
\xi(Q^2_0,Q^2) \; = \; \int^{Q^2}_{Q^2_0} \frac{dq^2}{q^2}
\frac{3\alpha_s(q^2)}{\pi} .
\end{displaymath}
In this kinematic regime GLAP effectively sums the \lq\lq double logarithms" in
$Q^2$ and $1/x$.  The double leading logarithm (DLL) form (15) shows that $xg$
grows, as $x \rightarrow 0$, faster than any power of log$(1/x)$ but slower
than
any power of $x$.  An example is the GRV set of partons \cite{GRV} which evolve
from a very low scale $Q^2_0 = 0.3$ GeV$^2$ so that the evolution length is
sufficiently long for the gluon to develop a steep DLL form in the HERA regime.
In a limited $(x,Q^2)$ region about $(\bar{x},\bar{Q}^2)$, the DLL form (15)
mimics a power law behaviour
\begin{displaymath}
xg(x,Q^2) \sim x^{-\lambda} ,
\end{displaymath}
with
\begin{equation}
\lambda \; = \; \left( \frac{36}{b_0} \frac{{\rm log}[{\rm
log}(\bar{Q}^2/\Lambda^2)/{\rm log}(Q^2_0/\Lambda^2)]}{{\rm log}(1/\bar{x})}
\right)^{\frac{1}{2}}
\end{equation}
where, for five flavours, $b_0 = 23$ and $\Lambda \approx 150$ MeV.  If we take
the GRV scale, $Q^2_0 = 0.3$ GeV$^2$, then in the lowest $x$ HERA regime
($\bar{x} \simeq 2 \times 10^{-4}$, $\bar{Q}^2 \simeq 10$ GeV$^2$) we find that
$\lambda \simeq 0.4$, which increases to $\lambda \simeq 0.48$ for $\bar{x}
\simeq 10^{-3}$, $\bar{Q}^2 \simeq 30$ GeV$^2$.   A higher starting scale
$Q^2_0$ will yield a lower effective value of $\lambda$.  For example if $Q^2_0
= 2$ GeV$^2$, as advocated in ref.\ \cite{KAID}, then in the above
$(\bar{x},\bar{Q}^2)$ regions we have $\lambda \simeq 0.24$ and 0.32
respectively.

Probably a better approach is to use a sufficiently high starting scale, $Q^2_0
\sim 5$ GeV$^2$, where perturbative QCD should be reliable, and to input a
singular gluon, $xg(x,Q^2_0) \sim x^{-\lambda}$ with $\lambda > 0$.  Then the
small $x$ behaviour is stable to evolution in $Q^2$, and overrides the double
leading logarithm form (15).  The MRS(D$^{\prime}_-$) and MRS(H) partons are
examples of this type of behaviour.

The GLAP-based approaches omit the summation of large log$(1/x)$ contributions
which are unaccompanied by large log$Q^2$ terms.  Thus at small $x$ and
moderate
$Q^2$ they can only be regarded as phenomenological parametrizations of the
gluon.  It is relevant to ask if they are sufficiently flexible to reproduce
the
behaviour expected from the log$(1/x)$ resummation.   Clearly there is a good
chance that the BFKL-motivated singular input form $xg \sim x^{-\lambda}$,
stable to evolution in $Q^2$, will suffice.

\vspace*{.5cm}

\noindent {\bf (b) BFKL expectations}

At small $x$ and moderate $Q^2 \, (\gapproxeq Q^2_0)$ we have only to sum the
large log$(1/x)$ contributions. (As before, we consider only the gluon as it is
the dominant parton at small $x$.) That is, at moderate $Q^2$ we must keep the
full $Q^2$ dependence, not just the leading log$Q^2$ terms.  The
strong-ordering
of the gluon transverse momenta, associated with the leading log$Q^2$
behaviour,
is therefore no longer appropriate, and we have to integrate over the full
$k_T$
phase space.  We thus have to work in terms of the unintegrated gluon
distribution $f(x,k^2_T)$  in which the \lq\lq last" $k^2_T$ integration along
the ladder (of Fig.\ 11(b)) is unfolded
\begin{equation}
xg(x,Q^2) \; = \; \int^{Q^2} \frac{dk^2_T}{k^2_T} f(x,k^2_T) .
\end{equation}
The leading $\alpha_s$log$(1/x)$ summation is accomplished by the BFKL (or
Lipatov) equation \cite{BFKL}, which may be written in the differential form
\begin{displaymath}
-x \frac{\partial f(x,k^2_T)}{\partial x} \; = \; \frac{3\alpha_s}{\pi} k^2_T
\int^{\infty}_0 \frac{dk^{\prime 2}_T}{k^{\prime 2}_T} \left[
\frac{f(x,k^{\prime 2}_T) - f(x,k^2_T)}{|k^{\prime 2}_T - k^2_T|} +
\frac{f(x,k^2_T)}{(4k^{\prime 4}_T + k^4_T)^{\frac{1}{2}}} \right]
\end{displaymath}
\begin{equation}
\equiv \; K \otimes f .
\end{equation}
 From (18) we see that the small $x$ behaviour of $f$ is controlled by the
largest eigenvalue $\lambda$ of the eigenfunction equation $K \otimes f_n =
\lambda_nf_n$, since  as $x \rightarrow 0$
\begin{equation}
f \sim {\rm exp} (\lambda {\rm log}(1/x)) \sim x^{-\lambda} .
\end{equation}

For fixed $\alpha_s$ the leading small $x$ behaviour of the solution of the
BFKL
equation, (18), has the analytic form
\begin{equation}
\frac{f(x,k^2_T)}{(k^2_T)^{\frac{1}{2}}} \sim
\frac{(x/x_0)^{-\lambda}}{[2\pi\lambda^{\prime\prime} {\rm
log}(x_0/x)]^{\frac{1}{2}}} {\rm exp} \left( \frac{-{\rm
log}^2(k^2_T/\bar{k}^2_T)}{2\lambda^{\prime\prime} {\rm log}(x_0/x)} \right)
\end{equation}
with
\begin{equation}
\lambda \; = \; \frac{3\alpha_s}{\pi} 4 {\rm log}2 \; \simeq \; 0.5,
\hspace*{1cm} \lambda^{\prime\prime} \; = \; \frac{3\alpha_s}{\pi} 28 \zeta (3)
\end{equation}
where the Riemann zeta function $\zeta (3) = 1.202$.  The characteristic
$x^{-\lambda}$ behaviour is the motivation of the $x^{-\frac{1}{2}}$ behaviour
assumed for $xg$ (and $x\bar{q})$ in set D$^{\prime}_-$ of partons.

A second feature of the solution (20) of the BFKL equation is the diffusion in
$k_T$ with decreasing $x$, as manifested by the Gaussian form in log$k^2_T$
with
a width which grows as (log$(1/x))^{\frac{1}{2}}$ as $x$ decreases.  The
physical origin of the diffusion is clear.   Since there is no strong-ordering
in $k_T$, there is a \lq\lq random walk" in $k_T$ as we proceed along the gluon
chain and hence evolution to smaller $x$ is accompanied by diffusion in $k_T$.
We foresee that the diffusion will be a problem in the applicability of the
BFKL
equation since, with decreasing $x$, it leads to an increasingly important
contribution from the infrared and ultraviolet regions of $k^2_T$ where the
equation is not expected to be valid.

Recently there have been many studies of the properties of the solutions
$f(x,k^2_T)$ of the BFKL equation \cite{LIP,AKMS}.  In some cases numerical
solutions have been obtained.

{\protect{\tenpoint
\begin{figure}[t]
\begin{center}
{}~ \epsfig{file=/pdg/rachel/adm/fig12.ps,width=200pt}
\end{center}
\caption{{\protect{\tenpoint The preliminary 1992-3 HERA data [31] compared
with the BFKL-based predictions of AKMS [30] and with expectations from MRS
[5, 15] and GRV [19] partons.}}
\label{12}
}
\end{figure}
}}

\noindent The usual technique is to solve (18) by
step-by-step integration down in $x$ from an input distribution $f(x_0,k^2_T)$,
at say $x_0 = 0.01$, determined from the gluon of one of the sets of partons.
Running $\alpha_s$ and shadowing effects have been incorporated, as well as a
study of different treatments of the infrared (and ultraviolet) regions, see,
for example, ref.\ \cite{AKMS}.  If we neglect shadowing effects, then, for a
given value of $k^2_T$, we find that the characteristic behaviour $f(x,k^2_T)
\sim Cx^{-\lambda}$ soon sets in with decreasing $x$.  The normalisation $C$ is
found to be much more sensitive to the infrared region than is the value of
$\lambda$.  Indeed for an infrared cut-off $k^2_0 \simeq O$(1 GeV$^2$) we find
$\lambda \simeq 0.5$, essentially independent of $k^2_T$ and only weakly
dependent on $k^2_0$.

\vspace*{.5cm}

\noindent {\bf (c) QCD predictions for $F_2$ at small $x$}

A representative set of QCD predictions and extrapolations for the small $x$
behaviour of $F_2$ at $Q^2 = 15$ GeV$^2$ is shown in Fig.\ 12, together with
the
most recent HERA data \cite{HERA}.  The curves are either predictions based on
the BFKL equation (AKMS) or phenomenological extrapolations of GLAP origin
(MRS,GRV).

The BFKL predictions are obtained \cite{AKMS} using the $k_T$ factorization
theorem
\begin{equation}
F_2(x,Q^2) \; = \; \int^1_x \frac{dx^{\prime}}{x^{\prime}} \int_{k^2_0}
\frac{dk^2_T}{k^4_T} f\left( \frac{x}{x^{\prime}},k^2_T \right) F^{(0)}_2
(x^{\prime},k^2_T,Q^2) + F^{{\rm bg}}_2(x,Q^2)
\end{equation}
where $x/x^{\prime}$ and $k_T$ are the longitudinal momentum fraction and
transverse momentum that are carried by the gluon which dissociates into the
$q\bar{q}$ pair, see Fig.\ 13(a).  The gluon distribution, $f$, is obtained by
solving the BFKL equation, (18).  The function $F^{(0)}_2$ corresponds to the
calculable quark box (and crossed box) amplitude for gluon-virtual photon
fusion, and $F^{{\rm bg}}_2$ is the slowly varying non-BFKL \lq\lq background"
contribution.  The value of $F_2$ is sensitive to the choice of the infrared
cut-off, $k^2_0$.  For the AKMS curve in Fig.\ 12, the value of the cut-off
($k^2_0 = 2$ GeV$^2$) is chosen so that there is a satisfactory description of
the $x > 10^{-3}$ data.  The $x$ shape or slope $\lambda =
-\partial$log$\bar{F}_2/\partial$log$x \simeq 0.5$ (where $\bar{F}_2 =
F_2-F^{{\rm bg}}_2)$ is, however, much less sensitive to the treatment of the
infrared region.  In other words the $x^{-\lambda}$ behaviour is generated by
BFKL dynamics with a determined value of $\lambda \simeq 0.5$, which is
essentially independent of $Q^2$ and of a reasonable variation of $k^2_0$.

Of course, strong shadowing corrections, if present, would change the small $x$
behaviour.  We know that parton shadowing (i.e. gluon recombination) will at
sufficiently small $x$ suppress the $x^{-\lambda}$ growth of the gluon density.
Conventional shadowing, arising from gluons spread uniformly across the proton
(of radius $R \sim 5$ GeV$^{-1}$), is found to give relatively little
suppression in the HERA region \cite{AKMS}.  However, if the gluons are
concentrated in \lq\lq hot-spots", say with $R = 2$ GeV$^{-1}$, then an
appreciable suppression occurs, see Fig.\ 12.  Again the normalisation is
adjusted by choosing the cut-off to be, in this case, $k^2_0 = 1.5$ GeV$^2$.

The other curves on Fig.\ 12, which are based on GLAP evolution, are much more
phenomenological.  Those denoted by MRS(D$^{\prime}_-$) and MRS(H) are obtained
from starting distributions which have $xg, xq_{{\rm sea}} \sim x^{-\lambda}$
with $\lambda = 0.5$ and 0.3 respectively.  Here $\lambda$ is a free parameter,
chosen at will.  In fact MRS(H), with $\lambda = 0.3$, was devised simply to
reproduce the HERA measurements of $F_2$  \cite{H1,ZEUS}.  The steep behaviour
of the GRV curve at small $x$ is generated by evolving from valence-like
starting distributions at a low scale $Q^2_0 = 0.3$ GeV$^2$;  the resulting DLL
form mimics $x^{-\lambda}$ with $\lambda \simeq$ 0.4 to 0.45 in the HERA
regime,
see (16).

Thus for GLAP evolution the steepness is either incorporated (as a factor
$x^{-\lambda}$) in the starting distributions or generated by evolution from a
low scale $Q^2_0$.  The steepness can be adjusted by varying $\lambda$ or
$Q^2_0$ so as to agree with the measurements of $F_2$.  On the other hand the
leading log$(1/x)$ BFKL prediction for the shape $F_2 \sim x^{-\lambda}$ with
$\lambda \approx 0.5$ is prescribed, but it remains to be seen how well it
survives a proper treatment of sub-leading effects -- their net effect is
expected to slow down the growth at small $x$.  The inclusion of $F^{{\rm
bg}}_2$ in (22) reduces the effective value of $\lambda$, and explains why BFKL
(AKMS) and MRS(H) with $\lambda = 0.3$ both give equally good descriptions of
the data.  Only at very small $x$, $x \sim 10^{-4}$, where $F^{{\rm bg}}_2$ is
relatively unimportant, does the BFKL-based prediction appear significantly
steeper than MRS(H) -- but even here shadowing effects could intervene.

It is also difficult to discriminate between BFKL- and GLAP-driven evolution
from the $Q^2$ dependence of $F_2$ at fixed $x$.  We expect a
$(Q^2)^{\frac{1}{2}}$ dependence from BFKL evolution (which applies at small
$x$
and moderate $Q^2$) as compared to the slower, approximately linear log$Q^2$,
behaviour characteristic of GLAP evolution.  In practice, at the accessible
values of $x$, subleading effects tend to slow the BFKL $Q^2$ evolution, so
that
it will be difficult to reduce the experimental systematic errors sufficiently
to identify the residual more rapid BFKL $Q^2$ behaviour \cite{AGKMS}.

\vspace*{1cm}

\noindent {\large {\bf 4.  Possible \lq\lq footprints" of BFKL dynamics}}
\vspace*{.5cm}

Although the observed small $x$ behaviour of $F_2$ is entirely consistent with
the $x^{-\lambda}$ BFKL perturbative QCD prediction, the above discussion shows
that the small $x$ data can be equally well described by phenomenological forms
evolved using conventional (next-to-leading order) GLAP dynamics.  The {\it
inclusive} nature of $F_2$, and the necessity to provide \lq\lq
non-perturbative" input distributions for the parton densities, prevents it
being a sensitive discriminator between BFKL and conventional dynamics.  For
this purpose it is necessary to look into the properties of the final state.

The two characteristic features of BFKL dynamics are the absence of
strong-ordering of the transverse momenta of the gluons along the chain and the
consequent
\begin{displaymath}
(x/x^{\prime})^{-\lambda} \hspace*{1cm} {\rm or} \hspace*{.5cm} {\rm
exp}(\lambda \Delta y)
\end{displaymath}
growth of the cross section, where $x$ and $x^{\prime}$ are the longitudinal
momentum fractions of the gluons at the ends of the chain, which spans the
rapidity interval $\Delta y = {\rm log}(x^{\prime}/x)$.  Recall that the
leading
$\alpha_s$log$(x^{\prime}/x)$ BFKL resummation gives $\lambda \approx 0.5$.
Some processes which exploit this behaviour are shown in Fig.\ 13.

\vspace*{.5cm}

\noindent {\bf (a) Deep inelastic events with a measured jet}

Mueller \cite{M} has emphasized the special features of deep inelastic
$(x,Q^2)$
events which contain a measured jet $(x_j,k^2_{Tj})$ in the kinematic regime
where (i) the transverse momentum of the jet satisfies $k^2_{Tj} \simeq Q^2$,
(ii) the jet longitudinal momentum, $x_j$, is as large as is experimentally
feasible $(x_j \sim 0.1)$, and (iii) $z = x/x_j$ is small.  A diagrammatic
representation of the process is shown in Fig.\ 13(a); for \lq\lq large" $x_j$
strong-ordering at the parton $a$-gluon vertex should be a good approximation.
The beauty of this measurement is that attention is focussed directly on the
BFKL $z^{-\lambda}$ type behaviour at small $z$, which arises from the
resummation of soft gluon emissions.  The choice $k^2_{Tj} \simeq Q^2$
neutralizes the ordinary gluon radiation which would have arisen from the
Altarelli-Parisi evolution in $Q^2$.  The differential structure function has a
leading small $z$ behaviour of the form \cite{DISJ}
\begin{equation}
x_j \frac{\partial F_2}{\partial x_j\partial k^2_{Tj}} \; \simeq \;
\alpha_s(Q^2) \left[ \sum_a x_jf_a(x_j,k^2_{Tj}) \right] z^{-\lambda} ,
\end{equation}
where the effective parton combination, $\sum f_a = g + \frac{4}{9}(q +
\bar{q})$, is evaluated at values of $x_j$ where the partons are well known
from
the global analyses.  This observable contains the anticipated $z^{-\lambda}$
behaviour, where, as before, $\lambda$ is the maximum eigenvalue of the BFKL
kernel, and so, in principle, it should allow an unambiguous identification of
$\lambda$. Another advantage of this process is that we can choose $k^2_{Tj}
\simeq Q^2$ sufficiently large so as to minimize the $k_T$ diffusion into the
infrared region.  Preliminary results indicate that the H1 collaboration
\cite{FELT} observe events at approximately the predicted rate.

{\protect{\tenpoint
\begin{figure}[t]
\begin{center}
{}~ \epsfig{file=/pdg/rachel/adm/fig13.ps,width=200pt}
\end{center}
\caption{{\protect{\tenpoint Diagrammatic representation of processes that may
be used to
identify BFKL dynamics: (a) deep inelastic scattering containing an identified
jet $a$, (b) $E_T$ flow accompanying deep inelastic scattering, (c) hadro- and
photo-production of dijets at large relative rapidity, $\Delta y$.  In the
leading log$(1/x)$ BFKL approximation $\lambda \simeq 0.5$.}}
\label{13}
}
\end{figure}
}}

\vspace*{.5cm}

\noindent {\bf (b) Transverse energy flow in deep inelastic events}

Due to the relaxation of strong-ordering of the gluon $k_T$'s along the chain
we
expect to find more transverse energy $E_T$ emitted in the central region
(between the current jet and the proton remnants) than would result from
conventional evolution, see Fig.\ 13(b).  For fixed $\alpha_s$ it is possible
to
derive \cite{KMSG} an analytic expression for the energy flow, $\partial
E_T/\partial$log$x_j$, in the photon-proton c.m.\ system.  The result is a
broad
Gaussian shape in log $x_j$ with a peak at $x_j \lapproxeq x^{\frac{1}{2}}$, a
width which grows as (log$(1/x))^{\frac{1}{2}}$ with decreasing Bjorken $x$,
and
a height which goes as $x^{-\lambda/2}$ and $(Q^2)^{\frac{1}{4}}$.  These
general features survive, at least qualitatively, a  full numerical treatment
--
quantitatively we find that the BFKL resummations yield \cite{KMSG} a fairly
flat central plateau in the HERA regime with $E_T \simeq 2$ GeV/unit of
rapidity, but much less $E_T$ if conventional dynamics are assumed.  No
hadronization effects have been included.

\vspace*{.5cm}

\noindent {\bf (c) Production of a pair of jets at large relative rapidity}

Recently there has been renewed interest in the original proposal of Mueller
and
Navelet \cite{MN} that the cross section for the production of a pair of
minijets should, according to the BFKL mechanism, rise roughly as exp($\lambda
\Delta y)$ as the rapidity interval $\Delta y$ between the jets becomes
increasingly large, see Fig.\ 13(c).  Quantitative studies \cite{PR}, at the
FNAL $\bar{p}p$ collider energy, show that the effect is masked by the fall-off
of the parton densities at large $x$ and by uncertainties associated with
sub-leading effects.  Rather it is proposed that the rate of weakening of the
azimuthal (\lq\lq back-to-back"-type) correlation between the jets, as the
rapidity interval increases, may be a better indicator of BFKL effects.

These studies have been extended \cite{GA} to the photoproduction of a pair of
minijets at HERA.  In this case minijets are expected to be dominantly produced
by \lq resolved' photon processes (i.e. processes mediated by the quark and
gluon constituents of the photon).  Thus again we have the diagrammatic
structure shown in Fig.\ 13(c).  The more limited interval of $\Delta y$  that
is accessible at HERA, in comparison to FNAL, may be partly offset by the
cleaner environment for minijet recognition.

\vspace*{.5cm}

\noindent {\bf (d) A compromise}

In general, the smaller the value of $x/x^{\prime}$ (or the larger the value of
$\Delta y$) the larger is the BFKL effect and the more dominant is the leading
log$(1/x)$ formalism.  As with all BFKL predictions, the reliability can only
be
quantified when the sub-leading corrections are known.  Moreover all the above
studies have been performed at the parton level and it will be necessary to
establish whether they survive full hadronization and detector simulation
effects.  For instance in the processes of Figs.\ 13(a) and (c) the jet $a$ has
to be resolved from the proton remnants and yet we want the jet to go as
forward
as possible to increase the $x/x^{\prime}$ (and $\Delta y$) \lq reach' of the
experiment.  If we increase the jet transverse momentum then the experimental
recognition is easier (and the infrared uncertainties in the BFKL predictions
are reduced), but the \lq reach' of $x/x^{\prime}$ (or $\Delta y)$ is decreased
and the leading BFKL effect is less dominant.

\vspace*{1cm}

\noindent {\large {\bf 5.  Conclusions}}
\vspace*{.5cm}

The parton densities of the proton are well determined in the region $0.02
\lapproxeq x \lapproxeq 0.7$ where high precision data for a wide range of
deep-inelastic and related processes are available.  The one combination,
$\bar{d}-\bar{u}$, that is not constrained by the deep-inelastic data, has for
the first time been determined at $x = 0.18$ by a measurement of the Drell-Yan
asymmetry in $pp$ and $pn$ collisions \cite{NA51}.  This measurement, together
with the $W^{\pm}$ rapidity asymmetry measurements \cite{BODEK}, probe fine
details of the quark densities.  It is therefore not surprising that the
existing MRS (and CTEQ) partons require minor modification.  The result is a
new
set of partons \cite{MRSA} -- MRS(A) -- which is essentially identical to
MRS(H)
at small $x$; recall that the latter parton set \cite{MRSH} was obtained by
incorporating the 1992 HERA measurements of $F_2$ \cite{H1,ZEUS} into the
analysis.  It is found that MRS(A,H) also give an excellent description of the
new preliminary 1993 higher statistics HERA data for $F_2$ \cite{HERA}.

The parton analyses are based on GLAP leading, and next-to-leading, log$Q^2$
evolution.  However, at sufficiently small $x$, the BFKL leading log$(1/x)$
resummation of soft gluons is more appropriate.  Then the predicted form for
$F_2$, found by AKMS \cite{AKMS}, is
\begin{equation}
F_2 \; = \; Cx^{-\lambda} + F_2^{{\rm bg}} ,
\end{equation}
where the normalisation $C$ of the BFKL contribution is dependent on the
treatment of the infrared region.  On the other hand the predicted value of
$\lambda$, $\lambda \simeq 0.5$, is much less sensitive to the infrared (and
ultraviolet) uncertainties in the solution of the BFKL equation.  It is a {\it
robust} prediction of (leading-order) BFKL dynamics and {\it not} a free
parameter.  The HERA $F_2$ data \cite{HERA} are well described by (24), where
the non-BFKL contribution $F_2^{{\rm bg}} \simeq F_2(x = 0.01) \simeq 0.4$.  In
fact the $F_2^{{\rm bg}}$ contribution is the reason why both AKMS (with
$\lambda$ dynamically determined to be 0.5) and MRS(A,H) (with $\lambda = 0.3$)
give equally good descriptions of the small $x$ data, despite their differing
values of $\lambda$.  It is remarkable that the data are consistent with the
precocious onset of leading-order BFKL dynamics, but clearly much work is
needed
to quantify the effect of the sub-leading corrections.

In the small $x$ HERA regime, GLAP evolution (from appropriately parametrized
starting distributions) is able to mimic the predicted BFKL behaviour.  Thus it
will be difficult to definitively identify BFKL dynamics from accurate
measurements of $F_2$.  Observables which are more sensitive discriminators
between BFKL and GLAP dynamics are deep-inelastic events containing an
identified forward jet, the transverse energy flow in deep-inelastic scattering
in the central region between the current jet and the proton remnants, and the
correlations between two jets that are produced at large relative rapidity.
Going from inclusive $F_2$ measurements to these new observables, in which we
have had to open up the final state, bring the problems of jet identification
and hadronization.  However the predicted effects are large and a  careful
experimental study should lead to an improved theoretical understanding of this
fascinating frontier of perturbative QCD.

\vspace*{1cm}

\noindent {\large {\bf Acknowledgements}}
\vspace*{.5cm}

It is a pleasure to thank Dick Roberts, James Stirling, Jan Kwieci\'{n}ski,
Peter Sutton, Adrian Askew and  Krzysztof Golec-Biernat for very enjoyable and
stimulating collaborations on the subject of this review, and especially Aharon
Levy for organizing such an excellent workshop.

\end{document}